# Synthesis and characterization of PEG-coated $Zn_{0.3}Mn_xFe_{2.7-x}O_4$ nanoparticles as the dual $T_1/T_2$-weighted MRI contrast agent

Bahareh Rezaei, Ahmad Kermanpur[*], Sheyda Labbaf

*Department of Materials Engineering, Isfahan University of Technology, Isfahan 84156-83111, Iran*

**Abstract**

Super-paramagnetic nanoparticles (NPs) have been widely explored as magnetic resonance imaging (MRI) contrast agents because of a combination of favorable magnetic properties, biocompability and ease of fabrication. MRI using traditional $T_1$- or $T_2$-weighted single mode contrast-enhanced techniques may yield inaccurate imaging results. In the present work, a $T_1/T_2$ dual mode contrast agent based on the super-paramagnetic zinc-manganese ferrite ($Zn_{0.3}Mn_xFe_{2.7-x}O_4$, x= 0, 0.25, 0.75 and 1) NPs with small core size and a hydrophilic PEG surface coating is reported. The TEM, TGA and FTIR results confirmed the formation of a uniform coating on the NPs surface. The MRI analysis revealed that the $Zn_{0.3}Mn_{0.5}Fe_{2.2}O_4$ NPs had the maximum image contrast compared to other zinc-manganese ferrite samples. Cell viability evaluations revealed that the coated and uncoated particles did not inhibit cell growth pattern. The present PEG-coated $Zn_{0.3}Mn_{0.5}Fe_{2.2}O_4$ NPs can be utilized as a suitable $T_1/T_2$-weighted MRI contrast agent for better diagnostic of abnormalities in the organs or tissues.

## 1. Introduction

The most potent and painless test that gives extremely clear images of the internal organs in the body is the magnetic resonance imaging (MRI) scan [1, 2]. Based on the magnetic relaxation processes of water protons on soft tissue of nearly every internal structure in the human body [1, 3-5], this method is a sort of diagnostic test that generates detailed images and functional information in a non-invasive and real-time monitoring manner [6, 7]. It is a distinguished device since there is no ionizing radiation during the imaging process and obviously reduces harmful side effects [2, 4, 8, 9]. However, this test typically provides poor anatomical details, and clinicians have some difficulties to distinguish between normal and abnormal tissues due to its low sensitivity [9, 10]. Hence, the clinical domains urgently require more reliable MR images. There is a potential to create more accurate and crisper images by adding contrast agents, which enables physicians to detect organs or *in-vivo* systems more clear. This opens up a wide range of MRI applications for therapeutic medicine in addition to diagnostic radiology. Despite the fact of shorter circulation time of $Gd^{3+}$ ions as a $T_1$-weighted MRI contrast agent, which renders them useless for high-resolution and/or targeted MRI [9, 11] and many concerns about potential trace deposition of Gd ions in the body, known as Nephrogenic Systemic Fibrosis (NSF) [12-14], which is a rare disease that frequently develops in patients with severe renal failure or after liver transplantation [15], Gd-based contrast agents can shorten the $T_1$ relaxation time effectively and provide brighter images in the regions of interest [16]. Following the increased awareness of this side effect, researchers have much more emphasis on alternative methods based on Mn-based complexes [15]. Although no scientific relationship has been proved between the NSF side effect and Mn so far, the metal is still known to pose some toxicity when inhaled. However, small amounts are essential to human health, but overexposure to free Mn ions may result in the neurodegenerative disorder known as 'Manganism' with symptoms similar Parkinson's disease [11].

---

[*] Corresponding author; Tel. (+98)3133915738; Fax (+98)3133912752; Email: ahmad_k@iut.ac.ir

Unlike $Gd^{3+}$ and $Mn^{2+}$ chelates, iron oxide nanoparticles (NPs) have achieved great attention due to the outstanding properties they exhibit at the nano-metric scale. A large number of benefits including biocompatibility, superparamagnetic behavior at room temperature, high saturation magnetization that can be tailored by size, shape, composition and assembly, tunable cellular uptake, biodispersibility, and large surface areas that make them a good candidate for polymer coating, conjugation with targeting molecules and other probes for achieving targeting and multimodal agents [17, 18] is reported for the iron oxide NPs. Super-paramagnetic NPs can be employed as $T_2$-weighted MRI contrast agents since they are more sensitive in the micro- or nano-molar range than Gd complexes [17]. Clinical MR imaging applications often use iron oxide-based NPs with strong magnetic moments as $T_2$-weighted MRI contrast agents. The limited usage of iron oxide NPs as $T_1$ contrast agents is due to their high transverse to longitudinal relaxivity ratio [19]. However, the use of superparamagnetic NPs in MRI is constrained by a negative contrast effect and magnetic susceptibility artifacts. Because the signal is frequently confused with signals from bleeding, calcification, or metal deposits and the susceptibility artifacts alter the background image, the resulting dark signal in $T_2$-weighted MRI may be exploited to mislead clinical diagnosis [18]. The $T_1$-weighted MRI contrast agents, however, have advantages over $T_2$-weighted MRI contrast agents. These advantages include better imaging quality, brighter images that can more effectively distinguish between normal and lesion tissues, and also the ability to provide better resolution for blood imaging. Nonetheless, in $T_1$- weighted MR imaging, some normal tissues (such as fatty tissue) may be mistaken for bright lesions that have been increased by $T_1$ contrast agents [20]. Therefore, efforts to integrate $T_1$ and $T_2$ imaging to prevent probable MRI artifacts and produce superior clinical images have been made as a result of the rising demand in the clinical diagnosis for both $T_1$- and $T_2$-weighted MR images. [18, 21]. Additionally, when several organ scans are required, injecting one dosage offers unmatched benefits to patients and doctors [16]. Super-paramagnetic NPs have the potential to exhibit significant dual $T_1/T_2$ relaxation performances when their sizes are decreased to less than 10 nm, according to some theoretical investigations [21-24]. Recently, super-paramagnetic iron oxide-gold composite NPs is synthesized by a green method [25]. It is shown that the NPs exhibited a high relaxivities ratio ($r_2/r_1$) of 13.20, indicating the potential as a $T_2$ contrast agent.

Surface modification is often practical to provide better stability under physiological conditions and prolong bloodstream circulation time, thereby increasing MR imaging quality [26]. This surface modification is known to restrict the uptake of plasma proteins (*i.e.*, corona proteins), which lowers the likelihood that macrophages will recognize and remove them [27]. In order to overcome the aforementioned difficulties, polymeric coatings on the surface of magnetic NPs are recommended [28]. In a recent work [29], iron oxide ferrofluid is synthesized by thermal decomposition using poly (maleic anhydride-alt-1-octadecene, noted as PMAO) as a phase transferring ligand. The results have demonstrated that the magnetic particles were fully covered at high coverage by long non-magnetic polymeric chains. It is shown that this ligand could improve the ferrofluid stability up to as long as 6 months. The MR images in solution and in rabbit using the prepared PMAO-coated magnetic NPs had the best contrast effect on $T_2$ weighted maps.

Polyethylene glycol (PEG) is a highly water soluble, hydrophilic, biocompatible, non-antigenic, and protein-resistant polymer that is easily eliminated through the kidneys and is not absorbed by humans' immune systems among all forms of polymeric coatings. PEG has also been frequently employed for linking anticancer medications to proteins to prolong their half-life, as well as for organ preservation [30. It also functions as an antibacterial, non-toxic lubricant and binder that is frequently used in a variety of medicinal applications [31, 32]. Additionally, PEG-capped magnetic NPs have demonstrated promise as effective and efficient magnetic hyperthermia candidates as well as multifunctional nano-carriers for the encapsulation of hydrophobic medicines [28]. In our previous work, we successfully synthesized $Zn_{0.3}Mn_xFe_{2.7-x}O_4$ (x=0, 0.25, 0.5, 0.75 and 1) NPs by a one-step citric acid-assistant hydrothermal method and reported the effect of citric acid concentration, pH of the medium and the amount of Mn addition on the structure, purity, and magnetic properties of the synthesized NPs [33]. According to the author's knowledge, citric acid-assistant hydrothermal synthesis of PEG-6000 coated $Zn_{0.3}Mn_{0.5}Fe_{2.2}O_4$ NPs as a dual mode $T_1/T_2$ imaging contrast agent have not been previously reported. In the present study, PEG surface coating is applied on the surface of the zinc-manganese ferrite NPs and then physiochemical properties of the optimized sample is thoroughly investigated. The mono-dispersed magnetic PEG-coated and uncoated Zn-Mn ferrite NPs containing different levels of Mn content is synthesized and the MR imaging of the NPs in the presence of external magnetic field is investigated.

## 2. Materials and Experimental Techniques

### 2.1. Materials

All raw materials, including $Fe(NO_3)_3 \cdot 9H_2O$, $NH_4OH$ 25%, $Zn(NO_3)_2 \cdot 4H_2O$, $Mn(NO_3)_2 \cdot 4H_2O$ and $C_6H_8O_7 \cdot H_2O$ (citric acid), $CH_3OH$, and PEG (MW=6000 g/mol) were purchased from Merck Co. with minimum purity of 99%.

### 2.2. Synthesis of Mn-Zn NPs

In order to synthesize $Zn_{0.3}Mn_xF_{2.7-x}O_4$ NPs, where x is the molar fraction of manganese ions ($Mn^{2+}$) from 0 to 1, various amounts of manganese iron nitrate, zinc nitrate and manganese nitrate were dissolved in 25 ml of distilled water. A reddish brown slurry was formed after adding a solution of 25% $NH_4OH$ which was added for the purpose of adjusting the pH of the media to 10. The resulting slurry was then washed with the deionized distilled water three times. Following the addition of the citric acid (CA), the mixture was rapidly stirred for 30 minutes before being placed to a 350 ml Teflon-lined autoclave with a 65% fill level. The autoclave was kept at 185 °C for 15 h and then cooled to room temperature [33]. Table 1 shows the experimental conditions of the synthesized samples. The uncoated samples were coded as NCZMX in which X is the molar fraction of $Mn^{2+}$ ions.

Table 1: The hydrothermal process parameters and the corresponding sample codes in the present work

| Sample code | Temperature (°C) | Time (h) | Citric acid (mmol) | pH | Molar fraction of $Mn^{2+}$(x) |
|---|---|---|---|---|---|
| NCZM | 185 | 15 | 3.5 | 10.5 | 0 |
| NCZM25 | 185 | 15 | 3.5 | 10 | 0.25 |
| NCZM50 | 185 | 15 | 3.5 | 10 | 0.5 |
| NCZM75 | 185 | 15 | 3.5 | 10 | 0.75 |
| NCZM100 | 185 | 15 | 3.5 | 10 | 1 |

## 2.3. Coating of Mn-Zn NPs

15 mg of NCZM50 and NCZM25 NPs were added to 1 ml deionized distilled water and then placed in an ultrasonic bath for 30 min. A polymeric solution containing 3 wt% PEG was dissolved in 1.5 ml of deionized distilled water and stirred for 30 min. The prepared magnetic ferro-fluid placed on a magnetic stirrer and then, the PEG solution were slowly added. This mixture was stirred at room temperature for another 1 h at ambient temperature (25 °C). Finally, the coated NPs were magnetically collected, washed with distilled water and dried in a vacuum oven at 40 °C for 24 h. The synthesized coated NPs are named as CZM25 and CZM50.

## 2.4. Cell viability

The MCF-7 cells were cultured in Dulbecco's modified Eagle's medium DMEM (Gibco 12800, UK) supplemented with 10% fetal bovine serum, 100 U/ml penicillin, 100 μg/ml streptomycin and 2 mM L-glutamine at 37 °C in a humidified atmosphere of 5% $CO_2$. The MG-63 osteoblast-like-cells were seeded at a density of 10,000 cells/well in a 96 well plate and cultured with complete medium containing NPs at concentrations of 50, 100 and 250 μg/ml. MCF-7 cells were exposed to particles for 24 h, after which Alamar Blue cytotoxicity assay was conducted and absorbance was measured at 450 nm using a micro-plate reader. The results represent the mean values ± SD of two individual experiments each performed in quadruplicate. Differences between groups were determined by student's t test with values of $p<0.05$ considered significant [34, 35].

## 2.5. Characterizations

Philips diffractometer, MPD-XPERT model, using CuKα radiation (λ = 1.5406 Å), was used for phase identification. Estimation of the average crystallite size (L) of the samples, using the full width at half maximum value (β) obtained from the spinel peaks located at every 2θ in the pattern, was carried out by the modified Scherer's formula. According to Scherer's modified formula, Lnβ (β in radians) is plotted against Ln(1/cosθ). A linear plot is obtained using the linear regression which is defined as Eq. (1). The intercept of the line would be Ln(kλ/L) (k=0.9); the value of L (mean crystallite size) can be obtained using all the peaks: [33, 36].

$$\mathbf{Ln\beta = Ln\left(\left(\frac{0.94\lambda}{L}\right) + Ln\left(\frac{1}{\cos\theta}\right)\right)} \quad (1)$$

The miller indices of the planes were extracted from the cards in the X'Pert software. Then, the mean lattice parameter was calculated based on Eq. (2) [37]:

$$a_i = d_i \times \sqrt{h_i^2 + k_i^2 + l_i^2}$$

(2)

The shape, size, and size distribution of NPs were investigated using transmission electron microscopy (TEM) with energy of 200 kV at Arya Rastak company in Tehran. A droplet of diluted magnetic flux was placed on a carbon coated copper mesh and placed at room temperature to allow water to evaporate. The average particle size of the produced zinc-manganese ferrite NPs from the TEM and SEM data was calculated by measuring the diameter of at least 100 NPs with ImageJ software. The data were fitted by a log-normal distribution curve and then the mean size was obtained.

Fourier transform infrared spectra (FTIR) were recorded in the range of 4000-400 $cm^{-1}$ to detect functional groups.

Saturation magnetization ($M_s$) values were conducted from the high field part of the measured magnetization curves, where the magnetization curve becomes linear and line's slope reaches to zero. Colloidal properties of the aqueous magnetic ferro-fluids were investigated using a Zeta Potential Estimator to measure the surface charge of NPs, hydrodynamic size, zeta potential and poly-dispersity index of NPs (in pH=7) under different conditions.

Thermo-gravimetric analysis (TGA) was used to investigate the presence of polymer coating on the surface of NPs.

MRI tests were performed with a 1.5 T clinical MRI instrument with a head coil working at 37 °C. For $T_1$ and $T_2$-weighted MRI of *in-vitro* cells at 1.5 T, the following parameters were adopted: [Mat (320*192), FoV (184*230), and TR (407)], [Mat (256*192), FoV (260*260), and TR (7)], [Mat (320*192), FoV (184*230), TR (2570)]. In order to simulate the physiological state of the body, PBS solution and water was used to create a positive and negative contrast in the images.

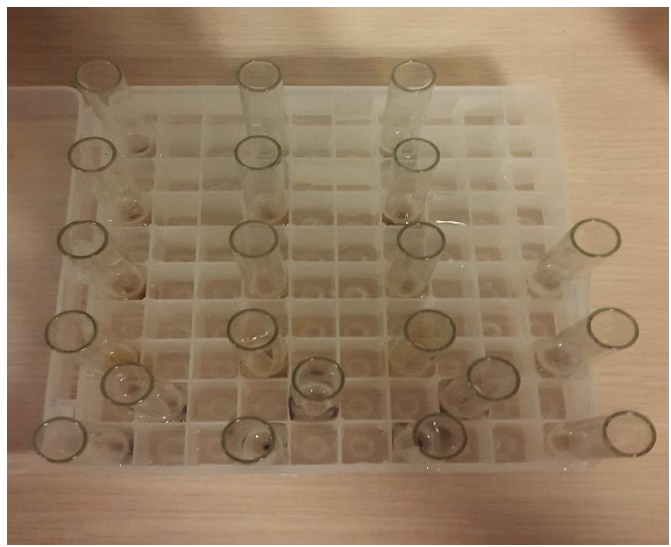

Fig. 1. Image of the prepared instrument for MRI imaging.

## 3. Results and Discussion

### 3.1. Structural properties

Fig. 2. shows XRD pattern of the NCZM50 NPs in which the diffraction peaks are in good agreement with planes (220), (311), (222), (400), (422), (511), (440), (620), (533) and (444) representing synthesis of pure spinel phase without the need for any calcination step. The crystallite size of the sample was estimated as 22 nm.

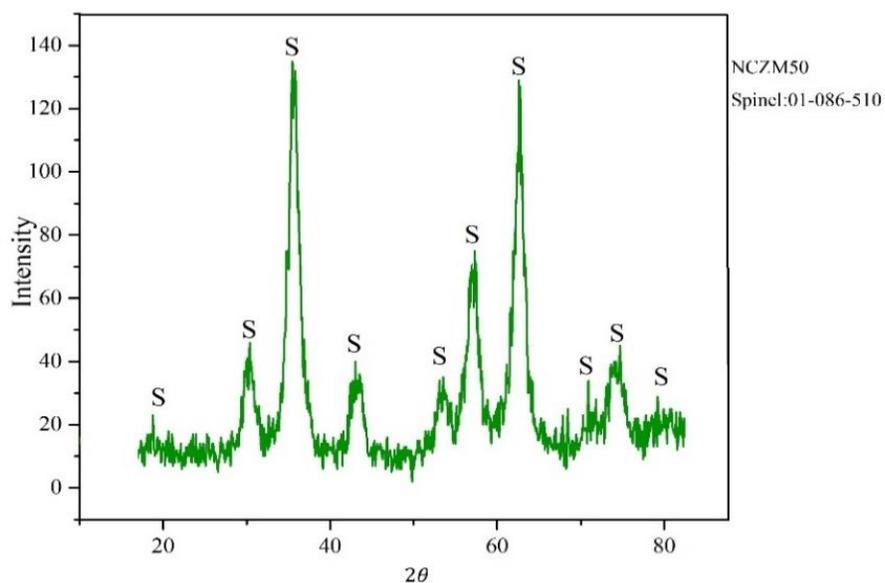

Fig. 2. The XRD pattern of the NCZM50 sample.

Surface coating is important in preventing NPs from agglomeration in physiological environment which also act as a barrier, effectively shielding the magnetic core against the attack of chemical species in the aqueous solution. Here, PEG was utilized to coat the optimized NPs. The FT-IR spectra of the pure NCZM50, the PEG-coated CZM50 NPs and the PEG are shown in Fig. 3. For the pure NPs, at around 3300 cm$^{-1}$, a strong wide band exists which is attributed to the O-H stretching vibrations of water molecules which are assigned to –OH group of CA absorbed by NCZM50 NPs (a structural bond). The stretching vibration of C-H corresponds to the peak at ~2925 cm$^{-1}$ [38, 39]. The absorption band at 1690-1760 cm$^{-1}$ is due to the vibration of asymmetric carboxyl group (-COOH) [28, 40]. Hence, it is suggested that CA binds to the NPs surface through carboxylate groups of citrate ions [28]. Furthermore, Fe-O stretching band as the characteristic peak of magnetite NPs was located at around 520 cm$^{-1}$ which is attributed to the Fe-O stretching vibration bond in tetrahedral sites and the absorption band in the 437 cm$^{-1}$ corresponds to a Fe-O vibrating bond in octahedral sites of ferrite phase [41]. Hydroxyl groups (-OH) of PEG are linked to the carboxyl group (-COOH) of citric acid (CA) for coating of $Zn_{0.3}Mn_{0.5}Fe_{2.2}O_4$ NPs. As it can be seen in Fig. 3, the highest peak for PEG curve showed a very small shift in PEG-coated sample. The peak at 1105 cm$^{-1}$ for pure PEG were shifted to lower frequencies which is a proof of C-O-C and C-O-H groups bonding with $Zn_{0.3}Mn_{0.5}Fe_{2.2}O_4$ NPs. The absorption band at 2884 cm$^{-1}$ can also be due to the H-C bonds stretching vibrations of the polymeric chain. The peaks corresponding to the bonds, C-H and C-O-C are the strong evidence to show that the synthesized magnetite NPs surface has been coated with PEG [38, 40].

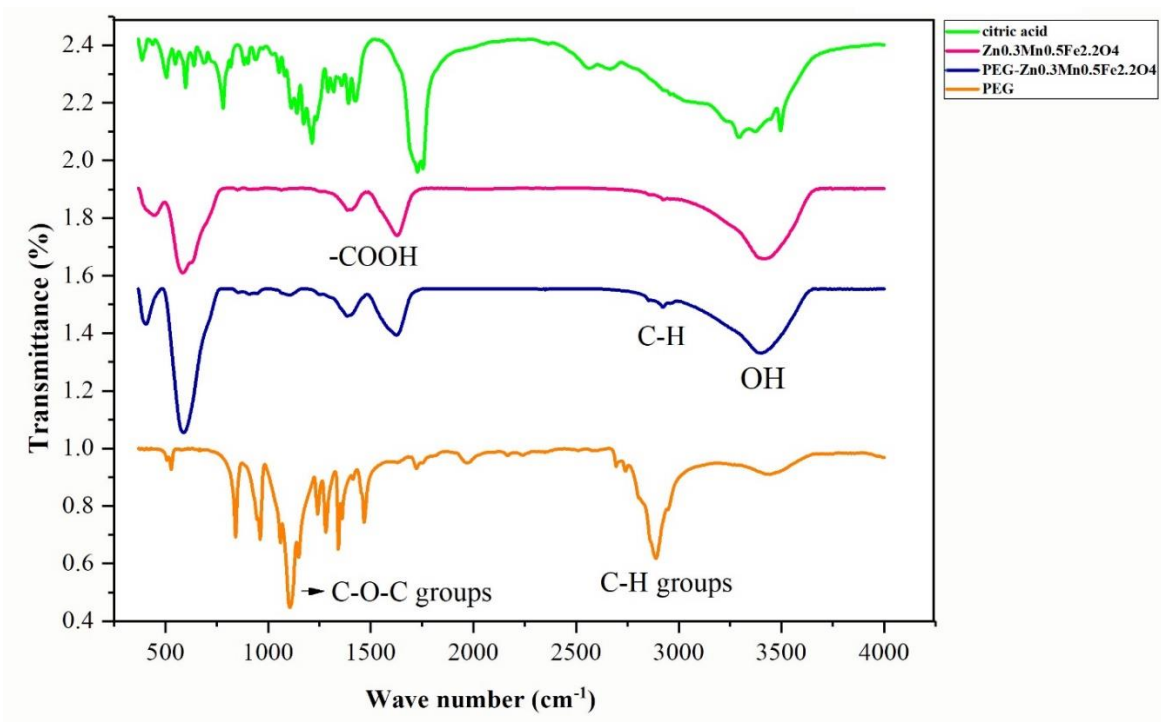

Fig. 3. The FT-IR spectra of the pure NCZM50 and PEG-coated CZM50 NPs along with the PEG coating and citric acid.

The presence of PEG layer on the NPs surface was also characterized by TGA which is presented in Fig. 4. The first stage of weight loss at a temperature about 32-35 °C can be related to the removal of water molecules (hydroxyl ions) that are physically absorbed to the surface of the NPs. This weight loss in the uncoated sample is 2.45% and in the coated sample is equal to 2.15%. The comparison of the first weight loss in the two samples shows that the total water loss of the NPs is more than coated NPs which is due to the total absence of water from the magnetic material structure [42]. The second step, starting at about 50-300 °C, results from the loss of organic groups that were conjugated to the surface of the particles. PEG desorption and subsequent evaporation were the causes of this weight loss. When 7.5 mg of PEG 6000 were used, the weight loss for particles was almost 24%, indicating 76% iron oxide in the polymer-coated NPs. Weight losses less than 15–20% can imply that the coverage of particle surface by the polymer is not complete [40].

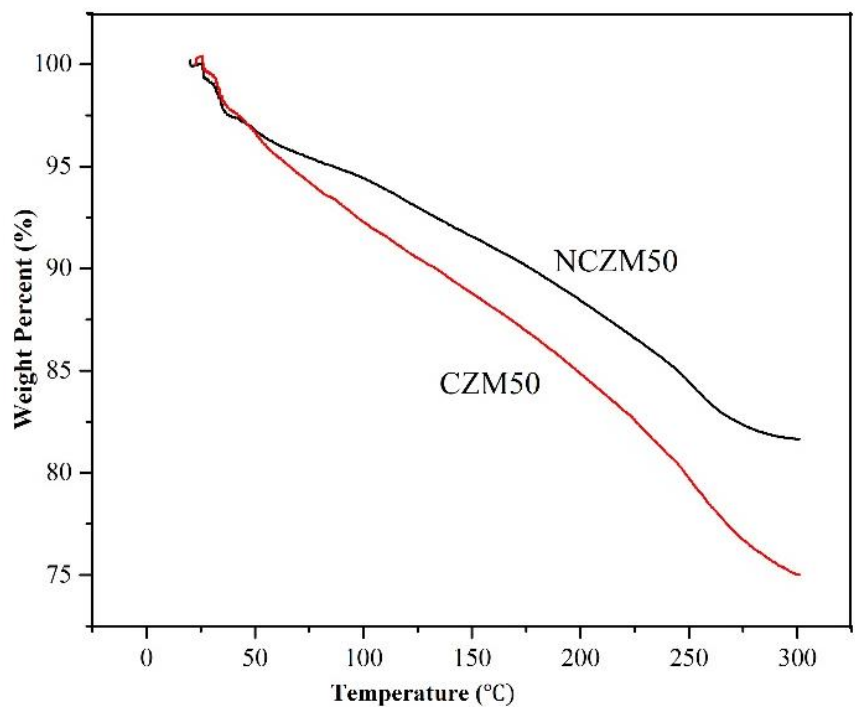

Fig. 4. The TGA result of the NCZM50 and CZM50 samples.

### 3.2. Microstructural analysis

Fig. 5 shows TEM micrograph and particle size distribution curve of the coated and uncoated samples. By using ImageJ software to measure the diameter of at least 100 NPs, the average particle size and the standard deviation was determined. As it can be seen, the synthesized NPs exhibit a rather uniform size distribution, shape, and morphology. The mean particle size of the coated NPs is a bit greater than that of the uncoated ones. It can be seen that NPs have become more dispersed after applying the coating in an aqueous medium. The average size of NPs obtained from the results of the TEM images before and after coating was 6.9±1.54 nm and 9.25±1.6 nm, respectively, which indicates that the polymer coating is applied on the surface of NPs at a low thickness [39].

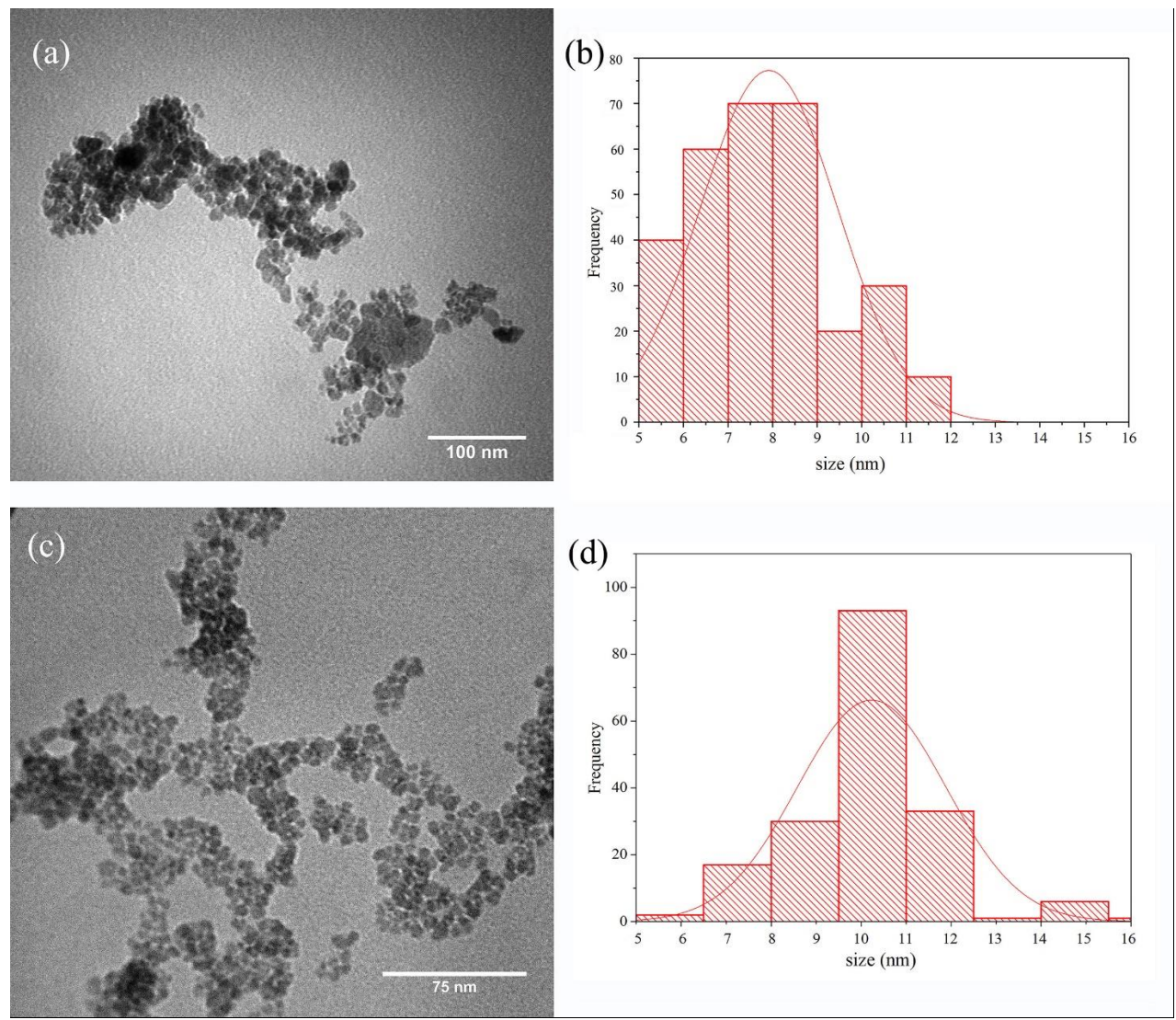

Fig. 5. (a, c) TEM images and (b, d) particle size distribution histogram of the (a, b) uncoated NCZM50 and (c, d) coated CZM50 NPs.

### 3.3. Stability and colloidal properties

The colloidal stability of magnetic fluids of NCZM50 sample was investigated using a zeta potential measurement at pH=7 and various time points. The result indicated that the NCZM50 NPs sample had a mean zeta potential of -48.86± 0.70 mV and a mean hydrodynamic size of 104 nm. According to the results, the strong negative charge of the NPs (caused by the presence of citrate ions on their surface) and the steric and electrostatic forces ensure their long-term stability in aqueous media [43].

Poly-dispersity index (PDI) of NCZM50 sample was found to be 0.306. PDI is a parameter for determining the particle size distribution of different NPs, which is obtained from photon correlation spectroscopic analysis. It is a dimensionless number calculated from the autocorrelation function and ranges from a value of 0.01 up to 0.7 for mono-dispersed and greater than 0.7 for poly-dispersed particles [44]. In general, the particle size between 10 and 100 nm have the longest circulation time; by contrast, it has been reported that particles of more than 200 nm tend to be immediately destroyed by one of the MPS organs [43] and tend to be eliminated by the RES [9, 45], those with diameters <10 nm are removed mainly by renal filtration, and particles larger than 400 nm (minimum diameter of capillaries) will be filtered by the lung [46].

### 3.4. Magnetic properties

The particle size and magnetization saturation values of different NPs are presented in Table 2. The room temperature M–H curve for NCZM50 and CZM50 samples is shown in Fig. 6. No hysteresis loop can be seen and the value of magnetization sharply increases with the external magnetic field strength. The M–H curve has an S-shape at the low field region, and the high field side of the curve is almost linear with the external field [47]. Saturation magnetization for the NCZM50 and CZM50 NPs is 55 emu/g and 38 emu/g respectively. The difference in particle size and the softening of the magnetization caused by the presence of PEG can both be used to explain this mismatch [38]. The magnetization curve of the CZM50 sample also revealed a negligible remnant magnetization at zero field, reflecting the super-paramagnetic behavior of the ferro-fluid. Since magnetic powder has a diameter much below the 20 nm cut-off expected for magnetite to show super-paramagnetic behavior, the lack of hysteresis at ambient temperature is consistent with this theory [48].

Table 2. The size and Ms values of different NPs

| Code | Chemistry | Size (nm) | Ms (emu/gr) |
|---|---|---|---|
| NCZM0 | $Zn_{0.3}Fe_{2.7}O_4$ | 14.5±2.7 | 47 |
| NCZM25 | $Zn_{0.3}Mn_{0.25}Fe_{2.45}O_4$ | 23.6±2.3 | 47 |
| NCZM50 | $Zn_{0.3}Mn_{0.5}Fe_{2.2}O_4$ | 6.9±1.5 | 55 |
| NCZM75 | $Zn_{0.3}Mn_{0.75}Fe_{1.95}O_4$ | 11.3±2.3 | 41 |
| NCZM100 | $Zn_{0.3}Mn_1Fe_{1.7}O_4$ | 6.7±2.4 | 37 |
| CZM50 | PEG coated- $Zn_{0.3}Mn_{0.5}Fe_{2.2}O_4$ | 9.3±1.6 | 38 |

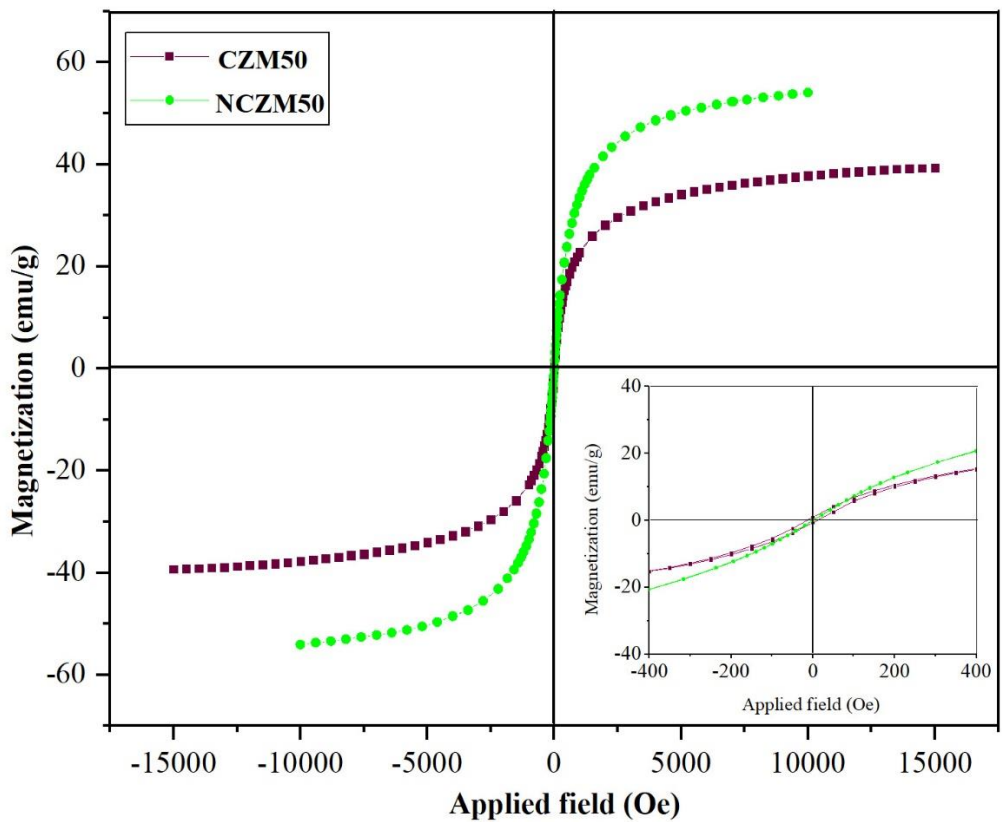

Fig. 6. The M vs H curves of the synthesized NCZM50 and CZM50 NPs.

**3.5. MRI analysis**

MRI examination of the body can be performed with several coil types, depending on the design of the MRI unit and the information required. Figs. 7(a-b) show $T_1$– and $T_2$-weighted MR images of $Fe_3O_4$ and Zn-Mn ferrite solutions recorded on a 1.5-T MRI scanner at room temperature at different concentrations (0.1, 0.15 and 0.2 mg/ml). As it can be seen, both $T_1$ and $T_2$-weighted MR images show a strong dependence of signal intensity on manganese concentrations and among the $Fe_3O_4$ control sample, Zn-based and Mn-Zn-based super-paramagnetic NPs, Mn-Zn ferrites represent a better MRI contrast [49]. This is due to the fact that $Mn^{2+}$ with five unpaired electrons, after $Gd^{3+}$, is the most powerful cation used as a MRI contrast agent [50]. Due to their greater paramagnetism and five unpaired electrons, divalent manganese ions ($Mn^{2+}$) have been shown to be a successful method of increasing the r1 of ultra-small iron oxide NPs. A peculiar mixed spinel structure, a greater saturation magnetization (Ms), and a high r2 of manganese doped iron oxide NPs result from the doped $Mn^{2+}$ with a higher magnetic moment (B=5.92) being able to fill both the tetrahedral (Td) and octahedral (Oh) sites in the crystal lattice. The doped $Mn^{2+}$ and ultra-small iron oxide NPs also exhibit synergetic enhancement, which will further enhance both r1 and r2 of Mn-iron oxide NPs, according to the embedding logic. The Mn-iron oxide NPs may therefore make superior candidates for dual-contrast CA [20]. Indeed, it has recently been discovered that decreasing iron oxide NPs below 10 nm improves their effectiveness as $T_1$ contrast agents, suggesting that this approach could be employed to create dual contrast agents. The utility of these NPs as $T_1$ contrast agents is unfortunately limited by the low $r_2/r_1$ values caused by the large decrease in r2 that occurred along with the increase in r1. To get over this restriction, alloy-based NPs which has a high Ms are a suitable candidate to achieve NPs with high MRI sensitivity [51]. The addition of $Mn^{2+}$ and $Zn^{2+}$ divalent cation ions to the spinel ferrite structure causes the mass magnetization of the material to rise, which enhances the magnetic characteristics. Therefore, the higher contrast in $Zn_{0.3}Mn_{0.5}Fe_{2.2}O4$ NPs with higher saturation magnetization can be justified [52]. As it is presented in Fig. 7, NCZM50 sample with core diameters about 6.7±1.54 nm and saturation magnetization about 55 emu/g is capable of producing dual positive and negative contrast in images [26, 53]. However, the length of the polymer chain, which relates to coating thickness, has a substantial impact on relaxivity as well. According to computer simulations, the physical exclusion of protons from the super-paramagnetic iron oxide magnetic field and the protons' residence period within the coating zone compete to decide the influence of coating thickness on relaxivity.

As it can be seen in the Fig. 8, the surface coatings also affect the relaxivity of NPs. Laconte et al. reported that the increased coating thickness would dramatically decrease the $r_2$ and $r_1$ relaxivity of mono-crystalline magnetic NPs. Therefore it is important to note that both the chemistry of coating and its thickness affect the value of $r_2$ and $r_1$ in which as the coating thickness increases, the ratio $r_2/r_1$ decreases. This is due to the inner hydrophobic layer excluding water, while the outer hydrophilic PEG layer allows water to diffuse within the coating zone [53] .

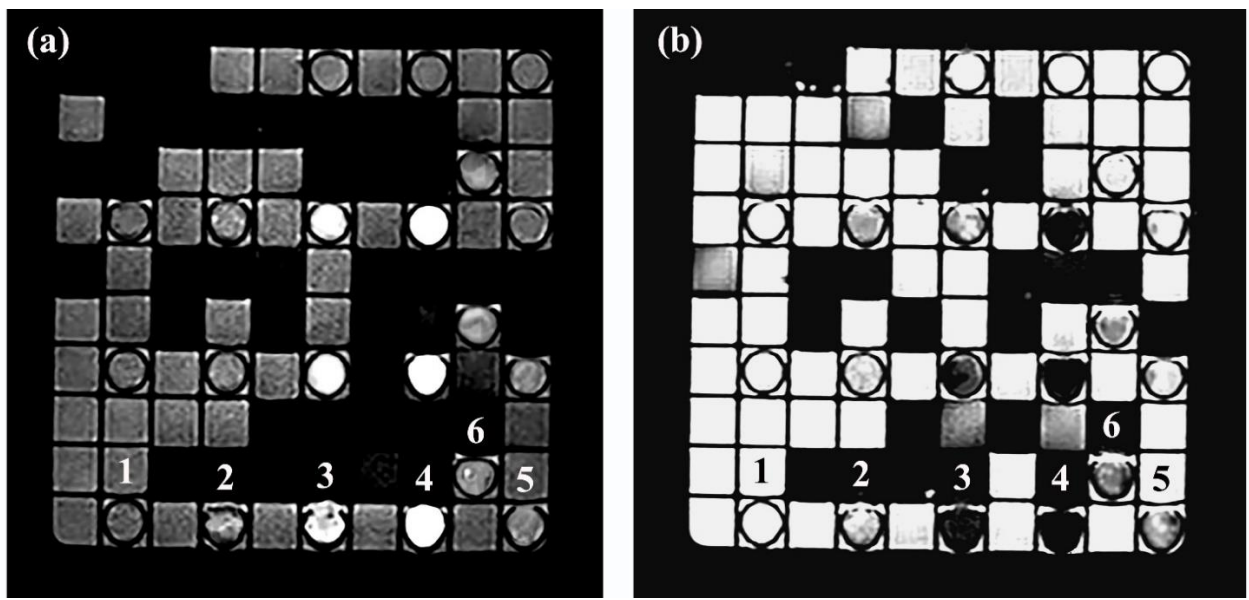

Fig. 7. (a) $T_1$-weighted and (b) $T_2$-weighted MR images of the uncoated $Fe_3O_4$ and Mn-Zn ferrite NPs at different concentrations indicated by different numbers: (1) $Fe_3O_4$ control sample, (2) NCZM0, (3) NCZM25, (4) NCZM50, (5) NCZM75, and (6) NCZM100.

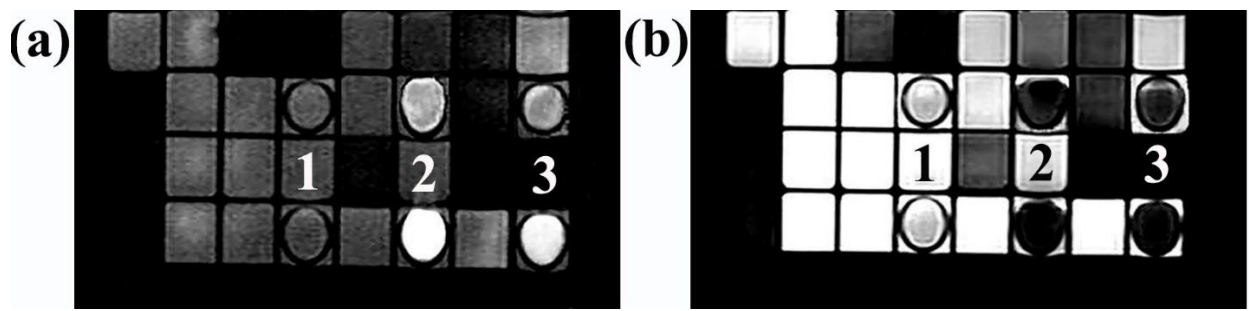

Fig. 8. (a) $T_1$-weighted and (b) $T_2$-weighted MR images of un-coated and uncoated samples indicated by different numbers: (1) $Fe_3O_4$ control sample, (2) NCZM50, and (3) CZM50.

### 3.6. Cell viability

Cytotoxicity evaluations of the uncoated and coated NPs were investigated by evaluating their cytotoxicity using MCF-7 cell line. The results of Alamar blue cytotoxicity assay are presented in Fig. 9. According to the results, a similar trend is observed in the activity of cells affected by different concentrations of NPs after 24 h compared with the control group. In general, coated and uncoated particles did not negatively change the cell growth process, and did not result significant reduction in cell viability. In fact, a better growth was observed in the presence of coated NPs.

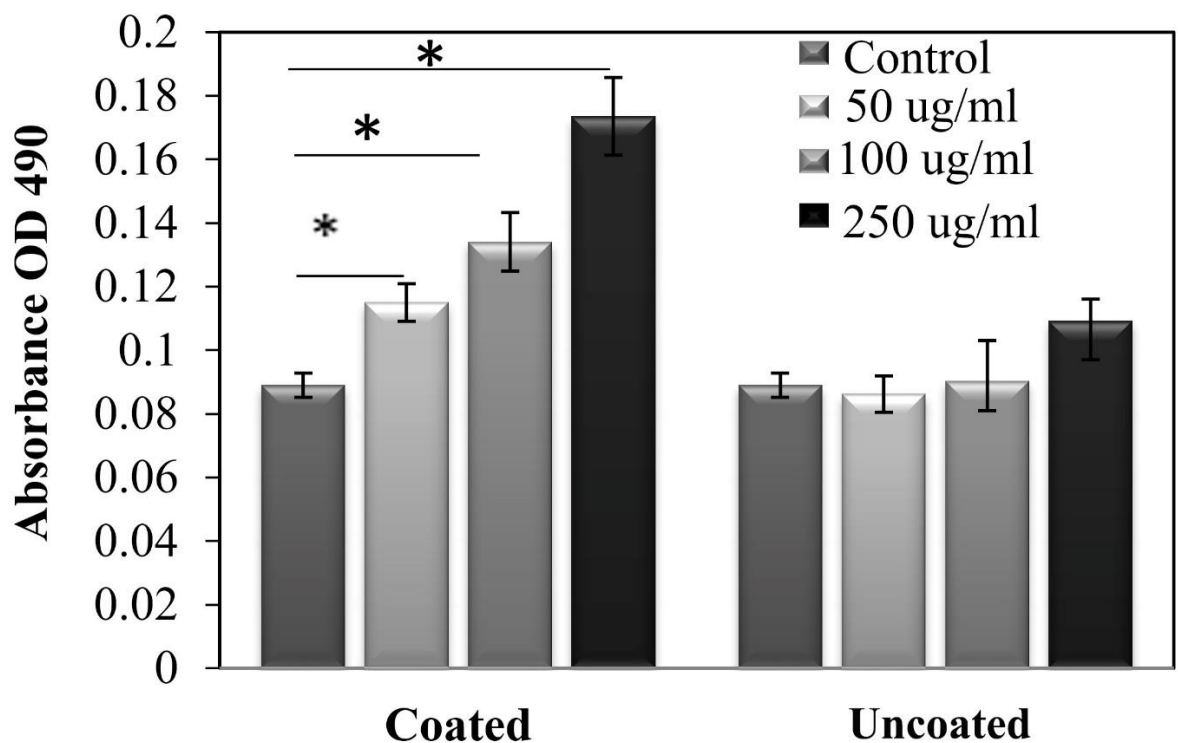

Fig. 9. The cytotoxicity assays performed on MCF-7 cells in the presence of coated and uncoated NPs after 24 h.

## 4. Conclusions

Mono-dispersed $Zn_{0.3}Mn_{0.5}Fe_{2.2}O_4$ NPs with an average size of about 6.9±1.5 nm were successfully synthesized by a facile, one step citric acid-assisted hydrothermal method. The NPs were stabilized with a layer of hydrophilic PEG and exhibited long-term colloidal stability in aqueous media at pH=7. The magnetic properties of the uncoated and coated Zn-Mn ferrite NPs were measured as 55 and 38 emu/g, respectively, showing super-paramagnetic behavior at room temperature. More significantly, the synthesized NPs displayed unexpectedly high $T_1$ and $T_2$ imaging contrast due to $Zn^{2+}$ and $Mn^{2+}$ doping and PEG-6000 coating. The present zinc manganese iron oxide NPs coated by PEG (ZnMnIONPs@PEG) are supposed to be a suitable candidate for application as $T_1/T_2$ dual contrast agent, as shown by *in-vitro* MR imaging. Interestingly, applying low thickness of PEG layer on the surface of the $Zn_{0.3}Mn_{0.5}Fe_{2.2}O_4$ NPs had no significant effect on the MR imaging.